\documentclass[a4paper]{jpconf}
\usepackage{graphicx}
\usepackage{xcolor}
\usepackage{soul}

\definecolor{tomcolor}{RGB}{0,127,255}

\begin{document}
\title{Progress on Optical Clock Technology for Operational Timescales}

\author{S.~Peil\textsuperscript{1}, W.~Tobias\textsuperscript{2}, J.~Whalen\textsuperscript{2}, B.~Hemingway\textsuperscript{1}, T.~G.~Akin\textsuperscript{1}}

\address{\textsuperscript{1} Precise Time Department, United States Naval Observatory, Washington, DC 20392, USA}

\address{\textsuperscript{2} Computational Physics Inc., Springfield, VA 22151, USA}

\ead{steven.e.peil.civ@us.navy.mil}

\begin{abstract}
While optical clock technology has advanced rapidly in recent years, incorporating the technology into operational timescales has progressed more slowly. The highest accuracy frequency standards for groundbreaking measurements do not easily translate to critical timing where continuous, uninterrupted operation over many months and years is required. For example, intermittent steering of a hydrogen maser with an optical standard fails to harness all of the dramatic improvements possible with optical technology. Here we present progress on development and integration of optical-clock technology for operational timescales.  An optical oscillator steered to an atomic fountain comprises a hybrid clock with optical-level stability at short times and a reliable long-term reference, and obviates the need for a steered maser. Atomic-beam optical clocks are being developed to support 24/7 operations at a level that improves upon the performance of the U.S. Naval Observatory's rubidium fountains.  An optical-lattice is being developed as a gold-standard frequency reference, complementing the role of the atomic beam clocks.
\end{abstract}

\section{Introduction - Operational Timescales}

Coordinated Universal Time (UTC) is the internationally agreed upon timescale derived from hundreds of atomic clocks maintained across the globe. Frequency standards are used to calibrate the timescale, imposing a fundamental time interval tied to the SI definition of the second and ensuring long-term stability through accurate evaluation of systematic perturbations. Clocks that operate continuously serve as a ``flywheel'', maintaining a timing output in between absolute frequency calibrations. Each metrology institute or timing lab maintains a number of clocks that contribute to UTC and also generate a local realization of UTC.

This general architecture for UTC and local timescales has been utilized since inception of atomic time a half-century ago. Perhaps the most significant evolution has been the contributing clock technology; thermal beam primary standards have been replaced by laser cooled atomic fountains~\cite{Wynands_2005}, and commercial cesium clocks are now weighted significantly less than hydrogen masers. In the past decade, rubidium fountains at the U.S. Naval Observatory (USNO) have contributed significant weight to UTC as continuous clocks (as opposed to secondary frequency standards), the first cold-atom clocks to contribute in this way~\cite{Peil_2016}.

The recent advent of optical-clock technology has led to the contribution of strontium (Sr) and ytterbium (Yb) optical lattices to UTC as (secondary) frequency standards, with anticipation that other optical-lattice or trapped-ion clocks will contribute in the near future.  On the other hand, the complexity of these systems prevents continuous, uninterrupted operation, and besides frequency calibration of a microwave clock (usually a maser), optical clocks have seen limited use in operational timescales.  There are currently no optical clocks contributing to UTC as flywheels, and only vapor-cell clocks~\cite{roslund2023optical, PhysRevApplied.9.014019} show significant promise as continuous clocks.  The limited use of optical clocks in timescales forces the continued use of microwave clocks and associated technology, limiting the timescale performance and capabilities over what is achievable with optical technology. We will discuss several efforts underway at USNO to more fully incorporate optical-clock technology into an operational timescale.

\section{Optical Oscillators} \label{oo}

The use of optical clocks for timing has almost exclusively consisted of an optical-lattice frequency standard intermittently measuring and steering a hydrogen maser~\cite{PTB_opt, NICT_month_opt} or ensemble of masers~\cite{NIST_opt}.  This architecture suffers from the lack of a suitable optical flywheel oscillator, relying instead on a proven microwave clock for continuity.  Short-term frequency stability is limited by the maser to $\sim 10^{-13}$ at 1~s, and the maser impacts longer-term performance to a degree that depends on the uptime of the lattice.

Optical oscillators, consisting of a cavity-stabilized laser and a frequency-comb divider, have matured significantly in recent years and are becoming much more reliable.  Fiber frequency combs routinely stay mode-locked for months on end, and the optical phase locks stabilizing the comb and the CW laser are even more robust. The radio-frequency signals generated maintain the frequency stability of the optical signal to a very high degree. While not yet reliable enough to replace masers in critical operations, optical oscillators are poised to find wider use in timing applications in the near future.

Working toward an experimental ``maserless'' timescale, we employ an optical oscillator constructed from telecom-wavelength components. A 1542~nm diode laser is stabilized to a high-finesse cavity made from a ULE spacer, silica mirror substrates and crystalline mirror coatings, resulting in a Hz-level linewidth. A fiber frequency comb centered at 1550~nm and with a 250~MHz repetition rate is phase-locked to the narrow laser and generates RF signals at 10~GHz and 200~MHz\footnote{The 4th harmonic of the repetition rate at 1~GHz is divided to generate 200~MHz.} via two different opto-electronic pathways. The system has proven to be robust, maintaining locks and producing an uninterrupted RF output for up to 6 months in a poorly regulated lab.

We further divide the 200~MHz signal to 5~MHz, the native frequency for many timing systems, and input the signal to our clock measurement system, which measures the scores of atomic clocks that contribute to the USNO timescale. While maintaining optical-clock-level stability when dividing to 5~MHz is not realistic~\cite{Hati_5M}, the 1~s instability is still well below that of a maser, and, for applications that can support a 200~MHz or higher frequency output, the impact of frequency division is negligible. 

The optical oscillator exhibits an average frequency drift of 3~kHz/day, with significant nonlinear variations strongly correlated with temperature over shorter times. Since optical frequency standards are not reliable enough for continuous operation, we steer the optical oscillator with a rubidium fountain. The USNO fountains have demonstrated reliable performance over 11 years of continuous operation and contribution to UTC, and an optical oscillator disciplined to an atomic fountain can form an operational clock combining the best of both systems.

The USNO rubidium fountains typically exhibit a white-frequency noise level of 1.5 to $2\times 10^{-13}$ when a quartz crystal drives the microwave frequency chain.  By instead using the 200~MHz signal generated from our optical oscillator, fountain performance can be improved, approaching more closely the quantum-projection noise limit that is as low as $5\times 10^{-14}$ in some of our systems. We average the frequency difference between the optical oscillator and fountain and feedback to the frequency-comb repetition rate to produce the steered output. 

\begin{figure}[h]
\centering
\includegraphics[width=30pc]{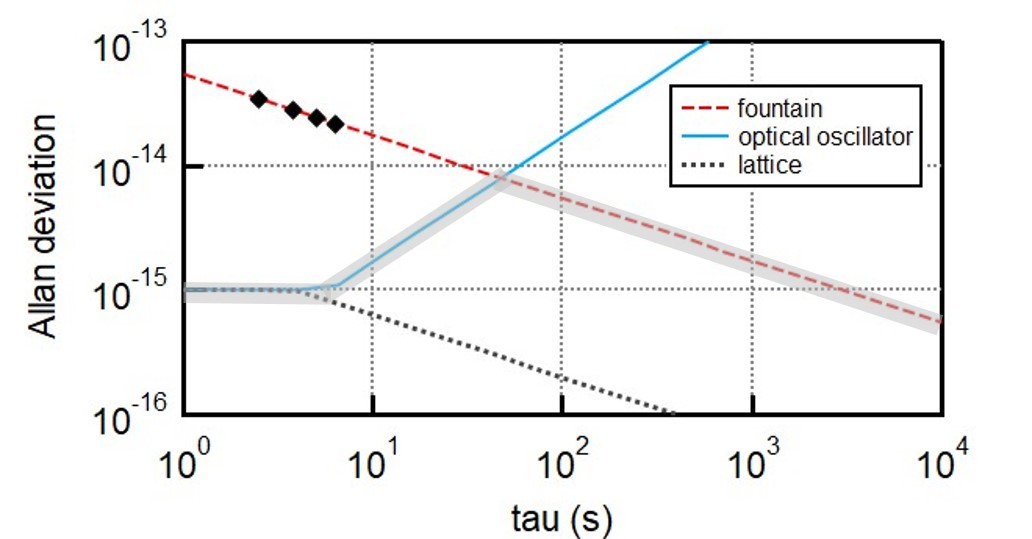}
\caption{\label{sigtau}Modeled behavior of hybrid optical-microwave clock. Solid and dashed curves labeled in graph legend show modeled Allan deviation for USNO rubidium fountain using optical oscillator for microwave generation (red, dashed), 10~GHz signal derived from optical oscillator (blue, solid), and lattice clock with $2\times 10^{-15}$ white-frequency noise level (black, dotted).  The thick grey band illustrates the expected performance of the composite clock, which should exhibit stability better than $10^{-14}$ for all averaging times.  The data points (diamonds) are measured in-loop fountain performance when running with the (unsteered) optical oscillator, validating the displayed white-frequency noise level.}
\end{figure}

This hybrid clock can be expected to exhibit an Allan deviation curve like that illustrated in Figure~\ref{sigtau}. Short-term stability governed by the local oscillator can be as low as $10^{-15}$ for the 10~GHz output, though degradation to $10^{-14}$ is expected when dividing to 5~MHz. The optical oscillator's frequency drift intersects the fountain's integration at about 40~s, after which the hybrid clock integrates as a fountain, shown with a white-frequency noise level of $6\times10^{-14}$. The frequency stability of the hybrid clock should be better than $10^{-14}$ for all averaging times, which would be a significant improvement over the short-term performance achievable with a maser-based timescale.  

A first test of this prototype hybrid clock was enabled by a portable Yb lattice developed at NIST, which made its maiden journey to USNO. The 10~GHz signal was measured against the 10~GHz output from the Yb lattice, which is expected to have comparable performance at 1~s but superior performance for all longer averaging times.  Multiple runs gave frequency records indicating 1~s instability of our hybrid clock at or below $5\times 10^{-15}$, and instability reaching $10^{-14}$ before integrating as white-frequency noise. There was no convenient, comparable reference to use to measure the stability of the 5~MHz signal.

This hybrid-clock demonstration shows the potential for implementing timescales without masers and associated steering synthesizers, harnessing the capabilities of optical technology for the timescale's front-end.





\section{Operational Optical Clocks}

The USNO rubidium fountains have been an integral part of the USNO timescale and a significant contributor to UTC for more than a decade.  Over that period, new timing requirements have emerged, and new applications, such as relativistic geodesy~\cite{Mehlstäubler_2018}, are becoming a reality.  Clocks that can perform significantly better than USNO's rubidium fountains, with 1~s stability of order $10^{-14}$ and a floor below $10^{-16}$, are expected to be needed to meet near future requirements.

With frequencies more than $10^4$ times higher than microwave transitions, optical clocks offer substantial headroom for improving upon a microwave standard. The natural or realized linewidth of the clock transition can be relaxed compared to the 1~Hz atomic fountain linewidth while still providing a significant improvement in stability.  
The $^{1}S_0 - {}^{3}P_1$ intercombination transition at 657~nm in neutral calcium has been of interest as an optical clock candidate off and on for decades~\cite{PhysRevLett.123.073202}. With a 400~Hz natural linewidth it is well suited for interrogation in an atomic beam, which offers a good compromise between the complexity of trapped atom systems and the non-ideal environment of vapor cells.

\subsection{Thermal}

The most robust architecture that we expect will meet our performance goals is Ramsey-Bord\'{e} spectroscopy on a thermal atomic beam.
We have studied this system extensively using a laser geometry that gives a fringe width (equivalent to the full width at half maximum) of 7.5~kHz, culminating in a stability budget that lays out the regulation requirements to reach our frequency stability floor~\cite{hemingway}.
New integrated optical assemblies will allow us to operate at a fringe width of 2.5~kHz, narrow enough to easily satisfy our short-term stability goals.

Excitation of the clock transition requires a cavity-stabilized laser, and we are investigating the tradeoffs of using a 657~nm laser and cavity versus an IR laser and cavity, with stability transferred to 657~nm via frequency comb. Detection using the $4s4p \hspace{1mm} {}^{3}P_1 - 4p^2 \hspace{1mm} {}^{3}P_0$ cycling transition at 431~nm provides an atom-background-free signal that enhances SNR at the expense of an additional laser at a challenging wavelength~\cite{Taylor:18}. 

The open question regarding a thermal-beam optical clock is the long-term frequency instability due to residual Doppler shifts driven by oven temperature instability (second-order Doppler) and by changes in optical alignment with respect to the atomic beam (first-order Doppler).  We are investigating using the technique of reversing the direction of laser-beam propagation (or ``$k-$reversal'') to compensate for frequency changes due to changing optical alignment~\cite{ITO}.  Figure~\ref{k_rev} shows Ramsey-Bord\'{e} fringes obtained from each direction of laser propagation in one of our atomic-beam systems.  Tracking the measured frequency for each direction should allow us to suppress the impact of alignment instabilities.

\subsection{Slowed}

Adding a stage of laser cooling to produce a slow calcium beam is a promising approach to significantly reducing residual Doppler instabilities.
We are investigating the tradeoffs involved in Ramsey-Bord\'{e} spectroscopy using a cooled calcium beam.

\begin{figure}[h] 
\begin{minipage}{16pc}
\includegraphics[width=16pc]{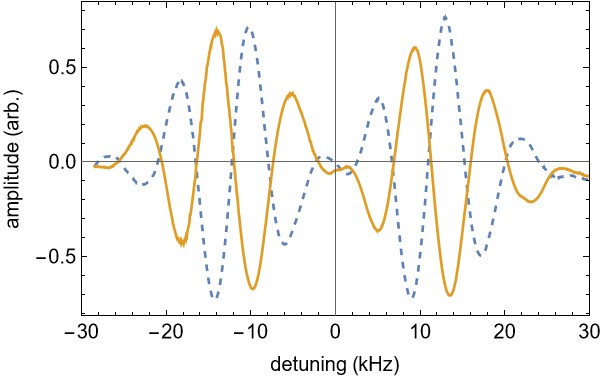}
\caption{\label{k_rev}Plot of Ramsey-Bord\'{e} signal for each direction of laser propagation. The 2.5~kHz fringe width enables clear resolution of the two recoil resonances separated by 22~kHz.  The horizontal ``detuning'' axis is roughly centered between these two resonances.}
\end{minipage} \hspace{1pc}
\begin{minipage}{16pc}
\includegraphics[width=16pc]{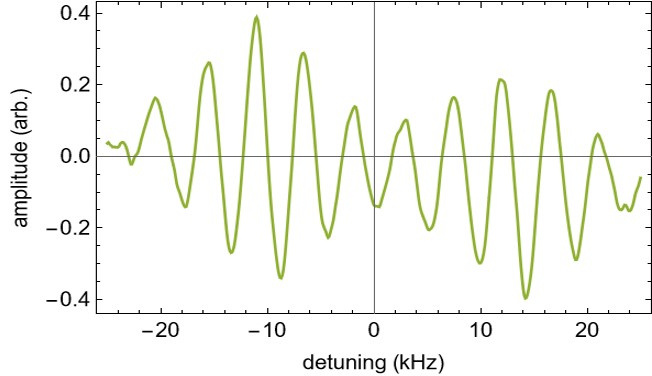}
\caption{\label{cold_ca}Ramsey-Bord\'{e} fringes measured in a slowed atomic beam. In a cold beam, the lifetime of the ${}^{3}P_1$ state reduces the contrast of the higher-frequency recoil resonance.}
\vspace{.8cm}
\end{minipage}
\end{figure}

Figure~\ref{cold_ca} shows Ramsey-Bord\'{e} fringes obtained with a 70~m/s calcium beam.  The cooled beam is generated using a longitudinal Zeeman slower and two 2D MOT stages for transverse cooling and beam steering.
Some of the 423~nm cooling light is used for detection.  The 2~kHz-wide fringes show higher coherence than in the thermal beam, with fringes extending throughout both recoil components.

Challenges in moving to a cooled beam include not just the additional complexity of the apparatus but also lower atom flux, impact of excited state (lifetime of 420~$\mu$s) decay, and narrower Fourier spectrum of the clock excitation (fewer transverse velocity classes addressed). Additional concerns include sensitivity of the cooled beam properties to the frequency and intensity of cooling light and potential light shifts from stray cooling light.

\section{Optical Frequency Standard}

The gold standard for an atomic frequency reference is an optical-lattice clock. The remarkable performance of such a system is rivaled only by its complexity - in addition to the frequency comb and high-finesse cavity needed for most optical clocks, a strontium (Sr) lattice clock also requires 7 CW lasers, some at challenging wavelengths and some at high powers.
Optical-lattice clocks have yet to demonstrate the continuous reliability required for timing operations, and their application to timescales to date have failed to harness their full potential.  

We are developing a Sr optical-lattice clock to improve timing capabilities at USNO. We add an additional CW laser to our architecture, a 1542~nm telecom laser that is stabilized to a high-finesse cavity and provides stability for the frequency comb and, in turn, all of the other CW lasers.  The addition of this laser provides the system with a robust flywheel based on telecom-wavelength components if one of the other 7 lasers, or some component of the atomic physics system, fails.
 
The conventional application of a Sr lattice to timing is to measure or steer the frequency of a hydrogen maser.  As discussed, this significantly limits the potential advantages of incorporating an optical clock into a timescale.  We are investigating the possibility of steering the optical oscillator to a rubidium fountain and creating a hybrid clock, as discussed in Section~\ref{oo}, with stability always better than $10^{-14}$. During periods when the Sr lattice is operational, the frequency stability will integrate along the lattice curve shown in Fig.~\ref{sigtau}, bringing the average instability for the composite clock system lower in proportion to the lattice uptime.  Additionally, we may choose to report the Sr lattice to the BIPM by calibrating the frequency of the rubidium fountain, rather than the frequency of a maser.  This could be beneficial depending on the averaging time required for the report.

\section*{Acknowledgments} 

This work is supported by Department of Navy Research, Development, Test and Evaluation funds provided for USNO’s Clock Development program. We are grateful to the NIST portable Yb lattice team, A. Ludlow, T. Bothwell and W. Brand.

\section*{References}

\bibliography{iopart-num}

\end{document}